\documentclass[aps,float,prd,psfig,amsmath,twocolumn]{revtex4-2}
\usepackage[dvipdfmx]{graphicx}
\usepackage{amsmath,amssymb,natbib,siunitx,booktabs,url}
\usepackage{color}
 \usepackage[normalem]{ulem}
\useunder{\uline}{\ul}{}

\newcommand{\lsim}{\lesssim}
\newcommand{\gsim}{\gtrsim}

\newcommand{\lmk}{\left(}
\newcommand{\rmk}{\right)}

\newcommand{\lkk}{\left[}
\newcommand{\rkk}{\right]}
\newcommand{\lla}{\left\langle}

\newcommand{\rra}{\right\rangle}
\newcommand{\so}{M_\odot}
\newcommand{\mch}{{\cal M}}

\newcommand{\beq}{\begin{equation}}
\newcommand{\beqa}{\begin{eqnarray}}
		  \newcommand{\eeq}{\end{equation}}
\newcommand{\eeqa}{\end{eqnarray}}

\newcommand{\pc}{{+,\times}}

\begin{document}

\title{Search for Higher Harmonic Signals  from Close White Dwarf Binaries in the mHz Band}

\author{Naoki Seto }
\affiliation{Department of Physics, Kyoto University, 
Kyoto 606-8502, Japan
}

\date{\today}

\begin{abstract}
Space-based gravitational wave (GW) detectors, such as LISA, are expected to
detect thousands of Galactic close white dwarf binaries emitting nearly
monochromatic GWs.
In this study, we demonstrate that LISA is reasonably likely to detect higher
harmonic GW signals, particularly the $(l, |m|) = (3, 3)$ mode, from a  limited sample of nearby close white dwarf binaries, 
even with small orbital velocities $v/c$ of 
  order $10^{-3}$.
The amplitudes of these post-Newtonian modes provide robust probes of mass
asymmetry in such systems, making them valuable observational targets, especially
in mass-transferring binaries. Long-term, coordinated detector operations will
further improve the prospects for detecting these informative signals.
\end{abstract}

\maketitle

\textit{Introduction}---
In the past ten years, LIGO and Virgo detectors have detected over 100 gravitational wave (GW) signals, mainly from binary black holes \cite{abbott2023gwtc}. Up until just before the mergers, these GW signals are dominated by the mode emitted at twice the orbital frequency, as predicted by the Newtonian quadrupole formula for a circular orbit \cite{blanchet2014gravitational,poisson2014gravity}.  In the so-called restricted phase approach, we focus exclusively on this dominant frequency component to facilitate an efficient data analysis \cite{cutler1994gravitational}. 

Even from circular binaries, higher harmonic (HH) signals are generated by the post-Newtonian (PN) effect, which is characterized by the PN parameter $\beta\sim v/c$ ($v$: the orbital velocity,  $c$: the speed of light) \cite{blanchet2014gravitational,poisson2014gravity}.  The amplitudes of the  leading order HH terms are suppressed by the factor 
$\beta$, relative to the Newtonian quadrupole term.  The LIGO-Virgo network has already detected HH signals from asymmetric systems, such as GW190412 and GW190814, in the strong gravity regime ($\beta \gtrsim 0.1$)  \cite{abbott2020gw190412, abbott2020gw190814}. These detections have significantly improved the estimation of binary parameters (e.g., spins and distances) and strengthened constraints on gravity theories.

Close white dwarf binaries (CWDBs) in our Galaxy are guaranteed GW sources for the future space interferometers such as  LISA \cite{amaro2023astrophysics}, Taiji \cite{hu2017taiji} and TianQin \cite{luo2016tianqin}. Indeed, above 4mHz, LISA is expected to detect $\sim 2000$ detached CWDBs  and a similar number of semi-detached CWDBs (AMCVn binaries) which transfer masses from the donors to the accreters \cite{Ruiter:2007xx,nissanke2012gravitational,Lamberts:2019nyk,amaro2023astrophysics,toubiana2024interacting}. Currently, only three Galactic CWDBs  have been identified in the frequency range above $4$ mHz through electromagnetic (EM) observations \cite{chakraborty2024expanding}. However, the number of identified CWDBs is expected to steadily increase as time-domain EM surveys continue to advance.

 Given the long radiation reaction timescale,   the orbital frequency of a CWDB remains nearly constant.  Its gradual frequency drift will be a basic observational target for LISA  \cite{amaro2023astrophysics}. 
 Importantly,  due to its small PN parameters $\beta=O(10^{-3})$ and the relatively large radii of white dwarfs, not only relativistic effects but also other astrophysical processes could  significantly influence the  frequency drift  of a CWDB. {For example, the dissipation rate of orbital energy may be temporarily enhanced by magnetic fields (e.g., \cite{Wolz:2020sqh,lau2024astrophysical}; see also \cite{2012MNRAS.421..426F} for tidal effects).
Moreover, an AM CVn binary can exhibit a negative chirp as a result of internal mass transfer \cite{amaro2023astrophysics}.
To better understand these complex systems, we should fully exploit the available GW data.}

The restricted phase approach appears to be a sufficient method for observing GWs from circular CWDBs. This is primarily due to the significantly smaller PN parameters of CWDBs,  compared to the binaries observed by LIGO and Virgo in the strong gravity regime. In fact, the detectability of HH modes in circular CWDBs has been overlooked until now.
 However, these PN modes are, in principle, sensitive to mass asymmetry. This characteristic plays a critical role, particularly in the long-term evolution and final outcomes of the binaries \cite{Nelemans:2001nr,amaro2023astrophysics}.
 
In this work, we demonstrate that LISA actually has a good chance to detect the HH modes for several nearby CWDBs. Due to their proximity to the Sun, these binaries will be bright GW sources and will be ideal systems for exploring weak and intriguing signatures not only the HH modes.   We also discuss the decomposition scheme for these signatures.



\textit{Quadrupole GW Emission}---
Let us consider a circular CWDB with an orbital frequency $f_{\rm o}$ and the component masses $m_a$ and $m_b$
 ($m_a\ge m_b$).  
 We define the total mass $m=m_a+m_b$ and the chirp mass $\mch=(m_am_b)^{3/5}m^{-1/5}$.
In addition, we define the mass difference ratio $\Delta ={(m_a-m_b)}/{m}$, 
which will later become important. 

For a circular orbit, 
the lowest quadrupole radiation is emitted only at the frequency $f_2\equiv 2 f_{\rm o}$. Its strain signal  is expressed as  
\beq
h_{+}^{[0]}=A H_{+}^{[0]},~~~h_{\times}^{[0]}=A H_{\times}^{[0]}
\eeq
in terms of the amplitude $A$ and the emission pattern functions $ H_{\pc}^{[0]}$.  In this work,  the numbers in  the square parentheses (e.g., $h_{+}^{[0]}$) show the PN orders and those in the subscripts (e.g., $f_2$) represent the wave frequencies in units of the orbital frequency $f_{\rm o}$.

The amplitude $A$  is given as
\beq
A\equiv  \frac{2\mch^{5/3}G^{5/3}\pi^{2/3}f_2^{2/3}} {c^4 d}
\eeq
with the gravitational constant $G$ and the distance $d$  to the CWDB.  
{Here, we apply the point-particle approximation and ignore the small correction of order \( O[(R/D)^5] \), which arises from the deformation of the stars (\(R\): stellar radius; \(D\): orbital separation) \cite{poisson2014gravity} (see also \cite{Broek:2012rs} for numerical evaluations showing that the corrections are at most \(\sim 1\%\)).
}

In the principle polarization axes, the functions  $H_{\pc}^{[0]}$ are given by 
\beqa
H_+^{[0]}&=&-(1+\cos^2 I)\cos2\Psi=a_{2+}(I)\cos2\Psi, \label{po1}\\
H_\times^{[0]}&=&-2\cos I \sin2\Psi=a_{2\times}(I)\sin2\Psi \label{po2}
\eeqa
with the inclination angle $I$ and the orbital phase $\Psi$ [satisfying $f_{\rm o}={\dot \Psi}/(2\pi)$] \cite{blanchet2014gravitational,poisson2014gravity}.
In Eqs. (\ref{po1}) and (\ref{po2}),  we defined the coefficients $a_{2+}$ and  $a_{2\times}$, which correspond to the $(l,|m|)=(2,2)$ modes of the spin-weighted spherical harmonics $_{-2}Y_{lm}(I,\Psi)$ \cite{blanchet2014gravitational}.


To characterize the total signal strength, we introduce the intensity function $E_2(I)$ (see appendix A for detail) by
\beq
E_2(I)\equiv \lkk a_{2+}(I)^2+a_{2\times }(I)^2 \rkk^{1/2}.  \label{e2}
\eeq

The matched filtering analysis is  the standard technique for detecting a regular GW signal characterized by a small number of parameters.  For the quadrupole GW from a circular CWDB, considering the response of detectors, the fitting parameters are the amplitude $A$, the sky position, the orientation angles (usually the inclination $I$ and the polarization angle), the initial phase and the GW frequency $f_2$ and its time derivatives.   

Due to the annual rotation of LISA's detector plane,  the sensitivity to linearly polarized incident GWs are efficiently averaged out (see e.g., \cite{yagi2011detector}). In fact,  in the relevant frequency regime,  if the integration period $T$ is a multiple of 1yr, the minimum sensitivity to the linearly polarized GWs  is estimated to reach
0.89 of their root mean square value \cite{yagi2011detector}. 
 Also counting the numerical factors, we can  estimate the signal-to-noise ratio $\rho_2$ of the quadrupole radiation as
\beq
\rho_2\simeq  \frac{A E_2T^{1/2}}{[S_n(f_2)]^{1/2}}\label{r2}
\eeq
with the effective noise spectrum $S_n(f_2)$ \cite{Robson:2018ifk}.

At present, from EM observations, we have only  {three} Galactic CWDBs at $f_2\ge 4$mHz \cite{chakraborty2024expanding}.  They are HM Cancri at $f_2=6.2$mHz \cite{strohmayer2021real,munday2023two}, ZTF J1539+5027 at $f_2=4.8$mHz \cite{burdge2019general}, and ZTF J0456+3843 at $f_2=4.2$mHz \cite{chakraborty2024expanding}.  In our study below, except for their distances, we use the first two binaries as representative systems.   Their basic parameters   are summarized  in Table I, including the Galactic coordinates $(l,b)$ with $b=0$ for the disk plane. Below, we briefly introduce the two    binaries. 

HM Cancri is an AMCVn binary, transferring mass from its donor ($m_b$) to its accreter ($m_a$), still in the inspiral phase ${\dot f}_{\rm o}>0$ \cite{strohmayer2021real,munday2023two}. The stability of the  mass transfer is considered to depend critically on the  asymmetry between $m_a$ and $m_b$.  \cite{Nelemans:2001nr}.  {The classical stability criterion (\( m_b/m_a \leq 2/3 \) \cite{1967AcA....17..287P}) corresponds to \( \Delta \geq 1/5 \).  
In contrast to the masses \( 0.55 + 0.27\,M_\odot \) (\( \Delta = 0.34 \)) presented in Table~I \cite{kupfer2018lisa}, the best-fit model in \cite{munday2023two} is \( 1.0 + 0.17\,M_\odot \) (\( \Delta = 0.71 \)), highlighting the uncertainties in mass estimation even with long-term, extensive EM observations.}
 
  Interestingly, at present, without a parallax measurement, the distance $d$ to HM Cancri  remains highly uncertain. From its observed proper motion together with a model for the transverse velocity distribution, Munday et al. \cite{munday2023two} presented a lower bound $d\ge 0.5$kpc.  The distance $d$ has been also estimated with various models for the observed EM emissions. However, partly due to the associated complicated accretion processes, the estimations have a large scatter (generally with $d> 0.5$kpc) \cite{munday2023two}.   

ZTF J1539 is a detached CWDB, showing a clear eclipsing light curve  \cite{burdge2019general}.  For eclipsing binaries, there exists a selection bias to detect nearly edge-on binaries  ($I\sim 90^\circ$).  From a spectroscopic analysis, Burdge et al. \cite{burdge2019general} estimated its distance $d=2.34\pm 0.14$\,kpc.  

In Table I, we present the signal-to-noise ratios $\rho_2$ for the quadrupole mode  estimated in \cite{kupfer2018lisa} and \cite{burdge2019general} for a 4yr operation of LISA.  Here   a tentative distance $d=5$kpc is used for HM Cancri. 

\begin{table*}[]
\caption{Basic parameters for HM Cancri  and ZTF J1539. }
\begin{tabular}{|c|c|c|c|c|c|c|c|c||c|c|c|c|c|}
\hline
          & $f_2 $ (mHz) & $(l,b)$       & $d$ (kpc) & $I$          & $m_1 (M_\odot)$ & $m_2 (M_\odot)$ & $E_2 $ & $\rho_2$ in 4yr & $\beta$ & $\Delta$  &$E_1$ &  $E_3$  \\ \hline
HM Cancri\footnote{Basic data from Kupfer et al. 2018 \cite{kupfer2018lisa}.} & 6.2          & (206.9,23.4) & 5.0  & $38^\circ$ & 0.55            & 0.27   & 2.26         & 211            &  0.0043& 0.34     & 0.57 & 1.57    \\ \hline
ZTF J1539\footnote{Basic data from  Burdge et al. 2019 \cite{burdge2019general}.}     & 4.8          & (80.8,50.6)  & 2.34           & $84^\circ$ & 0.61            & 0.21     & 1.03       & 143            & 0.0039 &0.49  & 0.63     & 1.15   \\ \hline
\end{tabular}
\end{table*}

\textit{post-Newtonian corrections}---
Next, we 
include the leading PN corrections.  The pattern functions $H_{+,\times}$ can be expressed as  \cite{blanchet2014gravitational,poisson2014gravity,kidder1995coalescing}
\beq
H_{\pc}=H_{\pc}^{[0]}+ \beta \Delta\,H_{\pc}^{[1/2]}+O(\beta^2).\label{wpn}
\eeq
Here the PN parameter $\beta$ is given by
\beq
\beta=\lmk  \frac{\pi Gmf_2}{c^3}  \rmk^{1/3}
=0.0043\lmk  \frac{m}{0.82M_\odot}\rmk^{1/3}\lmk \frac{f_2}{6.2{\rm mHz}}\rmk^{1/3}
\eeq
with the concrete values for HM Cancri.
This parameter   depends weakly on $m$ and $f_2$ for CWDBs in the LISA band. Meanwhile, we obtain a much larger value $\beta=0.4$ for $m=40\so$ and $f_2=100$Hz (typical values for binary black holes with the LVK network).

The pattern functions  $H^{[1/2]}_{+\times}$ are given as 
\beqa
H_+^{[1/2]}&=&-\frac{\sin I}8 [(5+\cos^2I)\cos\Psi-9 (1+\cos^2 I)\cos3\Psi]\nonumber \\
                    &=& a_{1+}(I)\cos\Psi+a_{3+}(I)\cos3\Psi  \label{hp05}\\
H_\times^{[1/2]}&=&-\frac34 \sin I \cos I ( \sin \Psi-3 \sin 3\Psi)\nonumber \\
                    &=& a_{1\times}(I)\sin\Psi+a_{3\times}(I)\sin3\Psi . \label{hc05}
\eeqa
{Here, we have used the point-mass approximation.  
These 0.5PN terms have the frequencies \( f_1 = f_{\rm o} \) and \( f_3 = 3f_{\rm o} \). The latter arises from the mass octupole mode, whose amplitude is  determined by the overall mass distribution of the binary \cite{maggiore2008gravitational}.  
The tidal effect can generate a correction of order \( O[(a/D)^5] \) to this mode, which is small, as expected from the case of the lowest quadrupole radiation \cite{Broek:2012rs}, supporting the validity of the point-mass approximation.  
Therefore, through the HH mode, we can robustly extract additional mass information about CWDBs.
 } 
 
 The patterns $a_{3+,\times}$ [defined in Eqs (\ref{hp05}) and (\ref{hc05})] are proportional to the $(l,|m|)=(3,3)$ components, while $a_{1+,\times}$  is given as a linear combination of the $(2,1)$ and (3,1) components \cite{blanchet2014gravitational}. 

As in Eq. (\ref{e2}), we define the intensity functions by
\beqa
E_1(I)&\equiv& \lkk a_{1+}(I)^2+a_{1\times }(I)^2 \rkk^{1/2}\label{e1}\\
~E_3(I)&\equiv &\lkk a_{3+}(I)^2+a_{3\times }(I)^2 \rkk^{1/2}\label{e3}
\eeqa
(see appendix A for detail).
In the last four columns in Table I, we presented the quantities related to the $f_1$ and $f_3$ modes.

As for  binary neutron stars (BNSs) with well measured masses, we   have the asymmetric ratios $\Delta$ much smaller than those in Table I (e.g. $\Delta=0.019$ for PSR B1913+16 and  $0.034$ for PSR J0737-3039).
Note also that  the number of Galactic BNSs is estimated to be much smaller than that of Galactic  CWDBs \cite{kyutoku2019detect,amaro2023astrophysics}. 

At the matched filtering analysis for the weak  $f_3$ and  $f_1$ modes, most of the involved parameters can be precisely determined   from the stronger $f_2$ mode. The truly independent parameter is the scaling factor $s\equiv \beta \Delta\propto m^{-2/3} (m_a-m_b)$, which determines the amplitudes of the two modes and  provides us with the information on the mass difference. Below, we denote $\rho_1$ and $\rho_3$ as the signal-to-noise ratios of the $f_1$ and $f_3$ modes, respectively.

As in the case for $\rho_2$ in Eq. (\ref{r2}), we can put
\beq
\rho_3\simeq \frac{A \beta \Delta  E_3T^{1/2}}{[S_n(f_3)]^{1/2}}
\propto \frac{m_a m_b (m_a-m_b)}{m}\frac{E_3(I)}{d}.
\eeq
From Eq. (\ref{r2}), we have 
\beq
\rho_3\simeq \rho_2 \beta  \Delta \frac{E_3(I)}{E_2(I)} \lmk  \frac{S_n(f_3)}{S_n(f_2)} \rmk^{-1/2}\simeq \rho_2  \beta \Delta  \frac{E_3(I)}{E_2(I)} \label{r3}.
\eeq

In the relevant frequency regime $2{\rm mHz}\le f\le 15$mHz, we approximately  have $S_n(f)\propto f^0$ for LISA \cite{Robson:2018ifk}.  This is because  the detector noise  is dominated by the optical path noise and the band is also  lower than the corner frequency 
$f_* =c/(2\pi L)=19$mHz ($L=2.5\times 10^6$km: LISA's armlength) \cite{Robson:2018ifk}. We thus put ${S_n(f_3)}/{S_n(f_2)}=1$ in Eq. (\ref{r3}). 

For the $f_1$ mode, we similarly obtain
\beq
\rho_1\simeq \rho_2  \beta\Delta \frac{E_1(I)}{E_2(I)}. 
\eeq
Here, given the detailed shape of LISA's noise curve $S_n(f)$ around $f\sim2$mHz \cite{Robson:2018ifk},
the  approximation  ${S_n(f_1)}/{S_n(f_2)}=1$ works less efficiently than ${S_n(f_3)}/{S_n(f_2)}=1$.  From $E_3(I)>E_1(I)$ (see Fig. 2), we should have $\rho_3>\rho_1$.

 In this work, for the two HH signals, we conservatively set the detection criteria  $\rho_{1,3}\ge5$. 
Unfortunately, from the numerical values presented in Table I, we obtain  $(\rho_{3},\rho_1)=( 0.21,0.078)$ for HM Cancri and  (0.31, 0.17) for ZTF J1539.

\begin{figure}
 \includegraphics[width=0.7\linewidth]{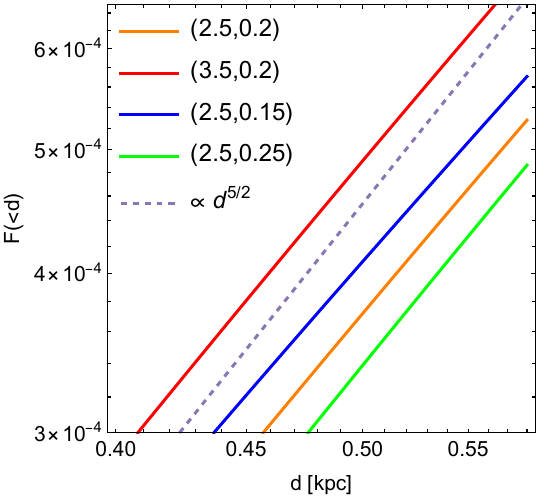} 
 \caption{The  cumulative fraction $F(<d)$ of Galactic CWDBs within distance $d$ from the Sun. We take four combinations $(R_D,z_D)$ for the disk scale lengths.  We assume the fraction $C_D=2/3$ for the disk component.
 }  \label{fig:del}
\end{figure}

\textit{Nearby binaries}---
As discussed earlier, the signal-to-noise ratios $\rho_3$ and $\rho_1$ are too small for HM Cancri and ZTF J1539 with their reference distances presented in Table I.  However, these two binaries are just the tip of the iceberg of the numerous CWDBs to be identified with LISA. Indeed, LISA is expected to detect  $N\sim 2000$ detached  CWDBs at $f_2\gsim 4$mHz, completing the whole Galaxy, and a similar number is estimated for AMCVn binaries \cite{nissanke2012gravitational,amaro2023astrophysics,toubiana2024interacting}. There could be undiscovered CWDBs much closer to us (e.g., buried  around the directions of the disk plane).   We thus evaluate the minimum distance, $d_{\min}$, of Galactic  CWDBs at $f_2\gsim4$mHz.

For the spatial distribution function of Galactic   CWDBs, we use the standard disk model given in the Galactic  cylindrical coordinate as
\beq
F(R,z)=\frac{C_D}{4\pi R_D^2 z_D} \exp\lmk {-\frac{R}{R_D}-\frac{|z|}{z_D}}\rmk \label{F}
\eeq
with  $C_D=2/3$ for the fraction of the disk component (ignoring the bulge component around the Sun) and the scaling lengths $(R_D,z_D)\sim({\rm 2.5,0.2})$kpc (see, e.g., \cite{nissanke2012gravitational}).   By integrating Eq. (\ref{F}) around the Sun at 
$(R,z)=(8.3,0)$kpc, we can evaluate the cumulative  fraction $F(<d)$ of CWDBs within a given distance $d$ from the Sun. In Fig. 1, we present the numerical results around $d\sim 0.5$kpc for various set of the scaling parameters $(R_D,z_D)$. 
We should obtain  $F(<d)\propto d^{3}$ at $d\ll z_D$ and $F(<d)\propto d^{2}$ at $ z_D\ll  d\lsim R_D$. Around $d\sim0.5$kpc,  we can approximately have $F(<d)\propto d^{2.5}$.

We can estimate the typical value $d_{\rm c}$ of the minimum distance $d_{\min}$,  by solving  $
N\times F(<d_{\rm c})=1
$ and obtain
\beq
d_{\rm c}\sim 0.5{\rm kpc}  \lmk \frac{N}{2000}  \rmk^{0.4} 
\eeq
with a relatively weak dependence on the total number $N$ and the scaling lengths.  In reality, the minimum value $d_{\min}$ has statistical fluctuation.  Using the fraction $F(<d)$, 
we can write down  a simple differential equation for the probability distribution function of $d_{\rm min}$ and readily obtain the  solution 
\beq
P(<d_{\min})=1-\exp\lkk -N\cdot F(<d_{\min}) \rkk \label{pdf}
\eeq
(also from basic expressions for the Poisson distribution).
By using Eq. (\ref{pdf}) and  the scaling relation $F(<d)\propto d^{2.5}$,  we  evaluate 
the 80\% interval of $d_{\min}$ as $[0.41d_{\rm c},1.4d_{\rm c}]$, with the median value $0.86d_{\rm c}$.

\begin{table}[]
\caption{Signal to noise ratios of LISA  for HM Cancri-like  and ZTF J1539-like binaries at $d=0.5$kpc in a 10yr integration. }
\begin{tabular}{|c|c|c|c|c|c|}
\hline
               & $f_2 $ (mHz) & $d$ (kpc) & $\rho_2$  & $\rho_3$  & $\rho_1$  \\ \hline
HM Cancri-like & 6.2    & 0.5     & 3340             & 3.38            & 1.24              \\ \hline
ZTF J1539-like       & 4.8     &0.5     & 1060              & 2.28            & 1.26         \\ \hline
\end{tabular}
\end{table}

In Table II, we present the signal-to-noise ratios $\rho_i$ ($i=2,1$ and 3) for  the two representative CWDBs now placed at the distance  $d=0.5{\rm kpc}\sim d_c$ with the extended operation period  $T=10$yr proposed for  LISA. Note that the maximum value $\rho_2\gsim 1000$ is generally consistent with previous estimations on CWDBs \cite{cornish2017galactic,littenberg2019binary}.

The magnitudes $\rho_3$ in Table II are close to the detection threshold 5. 
 {The distances for \( \rho_3 = 5 \) are respectively given by 0.338 kpc and 0.228 kpc. Using Eq.~(\ref{pdf}) and \( d_c = 0.5 \) kpc for the two model binaries, the probabilities of having them within the detectable distances become 31\% and 13\%.
} 
 Note that we can coherently combine the $f_3$ and $f_1$ modes. 
Then the relative accuracy of the scaling parameter $s$ is estimated to be $\delta s/s\sim 1/(\rho_3^2+\rho_1^2)^{1/2}$.

So far, 
we have discussed the HH signal search mainly  for LISA. 
 Around $f\sim 10$mHz, the design sensitivity $\sqrt{S_n(f)}$  of TianQin is close to that of LISA \cite{luo2016tianqin}, while Taiji is planned to have $\sim2$ times better sensitivity \cite{hu2017taiji}.  Therefore,   with their coherent signal integration for 10yr, we can further increase the ratios $\rho_i$ by a factor of $\sqrt{1^2+1^2+2^2}\sim 2.5$, indicating the importance of collaborative data analysis. {Indeed, after a simple analysis similar to the previous one for LISA alone, we now have the detection probabilities of 98\% and 75\% respectively for the two model binaries.
 }

\textit{Discussions}---
In this work, we have discussed GWs from circular binaries, including the leading PN correction.   However, even  at the Newtonian quadrupole order, an eccentric binary  emits GWs around multiples of   its orbital frequency $f_{\rm o}$. For a small  eccentricity $e\ll  1$, the amplitude of the $f_2$ mode is given as $A[1+O(e^2)]$ and those around the frequencies  $f_1$ and $f_3$ are $O(eA)$ \cite{Peters:1964zz} with the corresponding signal-to noise ratios $ O(e\rho_2)$. By measuring the amplitude of these two small signals,  we  can estimate  the eccentricity $e$ down to $e=O(1/\rho_2)$.  

Here, one might wonder whether we can distinguish the eccentricity induced Newtonian signal and the PN signal,  around the frequency $f_3$  (and similarly $f_1$).  {Actually, between the two signals, there is a small frequency gap given by the apsidal precession [e.g., due to the tidal and rotational deformations roughly given by \( f_2 (R/D)^5 \) as well as the relativistic correction \( \sim f_2 \beta^2 \)] \cite{seto2001proposal,willems2008probing,savalle2024detection}. For an observational period \( T \gtrsim 4 \) yr, the small frequency gap is likely to be resolved in the matched filtering analysis with the frequency resolution \( \sim T^{-1} \) \cite{seto2001proposal,willems2008probing,savalle2024detection}, allowing us to differentiate the PN and eccentricity-induced signals around the frequency \( f_3 \).
 }

Meanwhile, some gravitational theories predict the emission of anomalous polarization patterns  around the orbital frequency $f=f_1$ (e.g., dipole radiations)  \cite{Eardley:1973br,Will:1993hxu}. These signals could be interesting observational targets for the nearby GW sources and should be cleanly separated from our PN signals at $f=f_1$  with the ordinary transverse traceless polarization patterns.  Using the strong $f_2$ modes of the binaries,  we can precisely predict our PN signals, except for the scaling factor $s=\Delta \beta$.  Subsequently,   by taking an appropriate linear combination of the two TDI   data streams (relevant at $f\lsim f_*$),  we can effectively cancel  the contribution of our PN signals, independent of the factor $s$ (see, e.g., \cite{Chatziioannou:2012rf}). Then,  we can closely examine the anomalous polarization patterns at $f=f_1$.

A nearby CWDB would be a golden target also for follow-up EM observations.  LISA can localize the sky position of the binary  mainly from the Doppler modulation associated with its revolution  around the Sun.  For an observational period $T\gsim2$yr, the typical area of  the error ellipsoid in the sky is estimated to be  
\beq
\delta \Omega=(0.5{\rm arcmin})^2\lmk \frac{\rho_2}{3000} \rmk^{-2}\lmk \frac{f_2}{6.8{\rm mHz}} \rmk^{-2}  
\eeq
\cite{takahashi2002parameter}.
Unless the binary is almost at the face-on configuration (where $E_1\sim E_3\sim 0$),  using its GW data, we can predict  the expected periodic variations to be searched  in EM signals, such as the light curves or the spectral lines \cite{chakraborty2024expanding,burdge2019general,kupfer2018lisa}. 
   Considering the expected development on the time domain surveys, the binary might be detected  even before LISA's launch.

Once a binary is identified with an EM telescope, we can follow its long-term orbital phase evolution only with EM data (similar to the recent reports on the second derivative ${\ddot f}_{\rm o}$ for HM Cancri \cite{strohmayer2021real,munday2023two}), reducing the significance of the simple  phase analysis on GW data.  In contrast,  the amplitude measurement (including the HH search) can be done exclusively with GW measurement.

We might determine the distance $d$ to a nearby binary e.g., by an  EM parallax measurement. Then, from the GW amplitude   $A\propto \mch^{5/3}/d$ of the strong quadrupole mode, we can estimate the chirp mass $\mch$. Additionally measuring the  scaling parameter $s=\beta\Delta\propto (m_a-m_b)(m_a+m_b)^{-2/3}$ from the amplitudes of  the HH modes, we can, in principle,  solve the two masses $m_a$ and $m_b$ separately. 

The long-term future projects such as BBO \cite{Harry:2006fi} and DECIGO \cite{Kawamura:2011zz} have been proposed to explore the frequency range 10mHz-10Hz between LISA and ground-based detectors. Around 10mHz, these mid-band detectors have design sensitivities much better than that of LISA and could serve as powerful observatories for further studying the HH signals from Galactic CWDBs.

\textit{Summary}---
{We showed that space-based GW detectors, such as LISA, have great potential for detecting the HH modes of GWs for a handful of nearby CWDBs.  
} These overlooked signals are sensitive to the asymmetries of the binaries and would be critically important for complicated systems similar to HM Cancri. The detection prospects can be significantly enhanced through long-term and collaborative operations of the space GW detectors.

\acknowledgements
{The author would like to thank K. Kyutoku and T. Tanaka useful conversations.} 



\bibliography{ref}

\appendix
\section{Intensity functions}

\begin{figure}
 \includegraphics[width=0.7\linewidth]{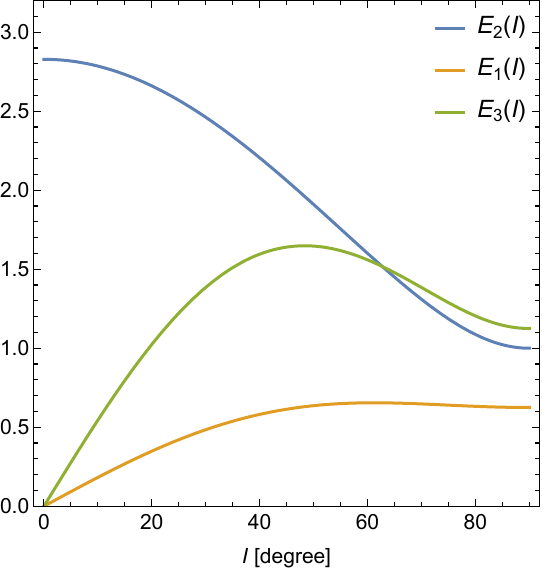} 
 \caption{The intensity  functions $E_2$, $E_1$ and $E_3$ defined in Eqs. (\ref{e2})(\ref{e1}) and (\ref{e3}).    
 }  \label{fig:del}
\end{figure}

In Fig. 2, we present the intensity  functions $E_2(I)$, $E_3(I)$ and $E_1(I)$ in the range $I\in [0^\circ,90^\circ]$.  The functions $E_2(I)$  monotonically decreases with $E_2(0^\circ)=2\sqrt2$ and $E_2(90^\circ)=1$.

 At the face-on configuration $I=0^\circ$, we have $E_1=E_3=0$, reflecting the spin-2 nature of gravitational radiation. They are not monotonic and respectively  take the maximum values  $E_1=(29\sqrt{29/2}-110)^{1/2}=0.65$ at $I=61.2^\circ$ and 
$E_3=(15\sqrt{5/2}-21)^{1/2}=1.65$ at $I=48.4^\circ$.
For reference, we provide the mean squared values 
$
\lla  E_2^2\rra_I={16}/5,~\lla  E_1^2\rra_I={5}/{14}$ and $\lla  E_3^2\rra_I={27}/{14}.
$

\end{document}